\documentstyle[12pt,openbib]{article}

\def\cao{\-\c c\~ao }

\def\coes{\-\c c\~oes }

\def\ao{\~ao }
\def\sao{s\ao }
\def\ii{\'{\i}}

\def\nao{n\ao }

\begin{document}

\begin{flushright}
\begin{tabular}{c c}
& {\normalsize  IFT-NI.001/98}\\
& {\normalsize  hep-ph/9807046 }\\
& \today
\end{tabular}
\end{flushright}

\begin{center}
{\bf\Large QUANDO UMA EXPERI\^ENCIA \'E CRUCIAL?}
\footnote{Apresentado no X
Col\'oquio de Hist\'oria da Ci\^encia, Campos do Jord\~ao, 8-12 de setembro de 
1997.},\footnote{Submetido \`a Revista Brasileira de Ensino de F\ii sica.}
\\
\end{center}
\vskip 2cm
{\center V. Pleitez \\
\vskip .8cm
Instituto de F\'\i sica Te\'orica\\
Universidade Estadual Paulista,\\ Rua Pamplona, 145\\
01405-900--S\~ao Paulo, SP, Brasil\\}
\vskip 1cm
{\center Resumo\\}
\vskip .5cm
Ainda que aceitemos que a F\'\i sica \'e, em \'ultima inst\^ancia, uma 
ci\^encia experimental, a rela\c c\~ao teoria-experimento est\'a longe de ser
trivial. Qualquer experi\^encia \'e sempre interpretada num determinado 
contexto te\'orico e, por sua vez, uma experi\^encia pode 
lan\c car novos desafios te\'oricos. Assim, n\ao podemos dizer
sem ambig\"uidade quando uma experi\^encia \'e crucial.

\newpage
{\center Abstract\\}
\vskip .5cm
Although we accept that Physics is, as a last resort, an experimental science,
the relationship between theory and experiment is far away from being trivial.
Any experiment is always explained within a determinate theoretical context
and, at the same time, an experiment can give suggestions
for theories or even can bring new theoretical challenges. Thus, we cannot say
without ambiguity when an experiment is a crucial one.

\newpage
\section{Introdu\c c\~ao}
\label{sec:intro}

A diferen\c ca fundamental entre a ci\^encia moderna p\'os  Francis 
Bacon, Galileo e outros, e a f\'\i sica em particular, com outras formas de 
estudar a natureza na antig\"uidade est\'a no papel que desempenha a 
experimenta\c c\~ao na verifica\cao ou falseamento das teorias~\cite{cohen}. 
Em \'ultima inst\^ancia a origem do m\'etodo cient\'\i fico encontra-se
na Gr\'ecia antiga por\'em, na chamada {\it Revolu\cao Cient\ii fica} do 
Sec.~XVII houve uma sistematiza\cao e aprofundamento do m\'etodo experimental,
isto \'e, a verifica\c c\~ao experimental de {\it algumas} das hip\'oteses 
b\'asicas de uma teoria em particular. 
Isto est\'a relacionado com o questionamento do
que \'e uma teoria. Por exemplo, uma teoria deve estar baseada 
apenas em quantidades observ\'aveis como queria Heisenberg? Em muitos casos 
\'e a teoria que diz o que deve e o que n\~ao pode ser observado. 
Esta era a vis\~ao de Einstein~\cite{wh1}.
A rela\c c\~ao teoria-experi\^encia, apesar da sua import\^ancia,
j\'a que afinal ela determina o que significa a ``verdade'' das leis da 
natureza, n\~ao est\'a bem definida.
As teorias representam o mundo real ou apenas as nossas percep\c c\~oes do 
mundo? A mesma pergunta vale para as experi\^encias.

Existe sempre uma diferen\c ca entre teoria e experimento. Por exemplo, nunca
verificaremos com exatid\~ao matem\'atica a lei da depend\^encia com $1/r^2$ 
da for\c ca
gravitacional.  
Qualquer resultado experimental confirmar\'a, ou n\~ao, esta 
lei com determinada precis\~ao. O fato de que o resultado seja {\it quase} 
igual a $1/r^2$ permite ``pular'' a dist\^ancia entre teoria e experimento e 
usar a matem\'atica para isso. (Mas as experi\^encias eliminam for\c cas
dominantes que variem como $1/r^3$ por exemplo.) 
Afinal, sem assumir que o expoente {\it \'e} ``2'' n\~ao
poder\'\i amos usar a an\'alise vetorial para sistematizar todos os efeitos 
observados e prever ou\-tros. A f\'acil matematiza\c c\~ao aparece como um
guia para aceitar uma teoria.~\footnote{Mas podemos especular se numa teoria 
estritamente {\it fractal} n\~ao seria poss\'\i vel fazer um tratamento de
uma lei de for\c cas que dependa de $1/r^{2+\alpha}$, sendo $\alpha$ um 
par\^ametro que dependa da dimens\~ao fractal do problema estudado. Claro na 
maioria dos casos $\alpha\approx0$.} Existem motivos mais profundos para 
considerar qualquer teoria como aproximada mas isto \nao ser\'a discutido aqui~
\cite{duhem},
por\'em vale lembrar as diferen\c cas entre Robert Boyle e 
Baruch Espinosa
~\cite{mchaui} ou Boyle e Thomas Hobbes~\cite{shapin}.
Segundo Boyle ``nas investiga\c c\~oes f\'\i sicas, frequentemente basta que 
nossas determina\c c\~oes cheguem muito pr\'oximas do fato, embora fiquem longe 
da exatid\~ao matem\'atica''.
J\'a Espinosa dizia ``res\-pon\-do que nunca precisamos da experi\^encia sen\~ao 
para aquilo que n\~ao pode ser conclu\'\i do da defini\c c\~ao de uma coisa''
~\cite{mchaui}. 

Assumiremos uma rela\c c\~ao teoria-experimento operacional, isto \'e,
n\ao vamos discutir o que \'e uma {\it teoria} e o que \'e uma {\it 
experi\^encia}. Usaremos a id\'eia intuitiva que todos temos a respeito desses 
conceitos. Aqui interessa-nos apenas discutir {\it a rela\cao} entre teoria e 
experi\^encia.
Por exemplo, uma experi\^encia crucial ({\it experimentum crucis}) sem
d\'uvida nenhuma \'e a experi\^encia dos dois prismas
de Newton. Antes do s\'eculo~XVII a cor dos corpos e a luz eram consideradas
dois assuntos diferentes. A cor por exemplo era considerada como uma 
propriedade material dos corpos. Descartes e Newton propuseram duas teorias
rivais sobre a natureza da luz e da cor. Para o primeiro, num contexto
mecanicista, a luz era a press\ao exercida por esferas de mat\'eria (das quais
o universo estava cheio) sendo a sensa\cao de cor, causada pelas diferentes
velocidades de rota\cao axiais das esferas. Nesta teoria o estado natural da luz
seria o branco e as cores apenas modifica\coes da ``brancura''. Para Newton,
a luz branca era uma mistura das diferentes cores. Na experi\^encia dos
dois prismas, na teoria de Descartes a refra\cao no segundo prisma deveria
implicar numa nova mudan\c ca das cores. Na teoria de Newton, por outro lado,
as cores do raio luminoso devem permanecer as mesmas. 
A experi\^encia de Newton decidiu a seu favor de maneira definitiva.
(Uma tradu\cao do trabalho de Newton encontra-se na Ref.~\cite{silva}).
\'E nesse sentido que consideramos uma experi\^encia crucial. Lembramos por\'em,
que no s\'eculo~XVII a quest\ao n\ao foi aceita de maneira 
imediata~\cite{shapin}.

Assim, podemos 
perguntar: Quando \'e possivel dizer se uma experi\^encia foi {\it crucial} ? 
Ou seja quando ela permitiu selecionar a teoria certa entre duas ou mais 
teorias rivais como no caso da experi\^encia de Newton mencionada acima, ou 
mesmo que confirmou ou negou uma determinada teoria. 
Vamos ver que n\ao existe uma resposta \'unica para esta quest\~ao. Um \'unico
contra exemplo pode derrubar uma vis\ao geral sobre a rela\cao 
teoria-experimento. Podemos encontrar exemplos nos quais
\begin{description}
\item[{\it i)}] uma experi\^encia foi realizada antes do que seu conte\'udo te\'orico
estivesse desenvolvido e, por isso, n\~ao foi reconhecida como crucial na
\'epoca.
\item[{\it ii)}] experi\^encias para as quais os preconceitos te\'oricos adiantaram a 
sua validade.
\item[{\it iii)}] experi\^encias corretas que foram interpretadas em contextos 
te\'oricos errados.
\item[{\it iv})] experi\^encias que n\~ao foram guiadas nem teorica nem 
epistemologicamente.
\end{description}

Esta lista n\~ao esgota o estudo da rela\c c\~ao teoria-experimento mas ilustra 
muito bem que sua rela\c c\~ao n\~ao \'e trivial. Experi\^encias do tipo do 
\'\ item {\it iv)} n\~ao ser\~ao discutidas aqui em detalhe. 
Apenas mencionamos entre esse tipo de experimentos: os raios-X e a 
radioatividade natural (talvez a descoberta da viola\cao da simetria 
discreta $CP$ seja um deste tipo).
 
A seguir, consideraremos alguns exemplos dos tr\^es tipos de experi\^encias
descritos nos itens {\it i)--iii)} acima. Na Se\c c\~ao \ref{sec:sec2} 
consideramos o caso da viola\c c\~ao da paridade nas intera\c c\~oes fracas como 
exemplo do tipo {\it i)}. Na Se\c c\~ao~\ref{sec:sec3} consideramos quatro 
experi\^encias que exemplificam a situa\c c\~ao {\it ii)}. O caso {\it iii)} \'e 
analisado em duas experi\^encias na se\c c\~ao~\ref{sec:sec4}. Uma \^enfase maior 
\'e dada \`as experi\^encias da viola\c c\~ao da paridade e \`a de Stern-Gerlach 
na Se\c c\~ao~\ref{sec:5}. As conclus\~oes encontram-se na \'ultima Se\c c\~ao.

\section{Experi\^encias n\~ao reconhecidas como \\ cruciais na \'epoca}
\label{sec:sec2}
No primeiro tipo de experi\^encias podemos mencionar o descobrimento da 
vi\-o\-la\-\c c\~ao da simetria sob {\it invers\~oes espaciais} ou 
{\it paridade} (ou seja a transforma\c c\~ao das coordenadas espaciais que faz
$\vec{x}\to-\vec{x}$) nas intera\c c\~oes fracas. Pode-se mostrar que esta
transforma\cao \'e equivalente a uma reflex\ao especular {\it i.e.},
a que acontece quando nos olhamos num espelho. Do ponto de vista 
te\'orico a possibilidade deste efeito foi colocada, n\~ao pela primeira 
vez, mas sim de maneira decisiva, por Lee e Yang em 1956. 

Em 1957 tr\^es experi\^encias diferentes confirmaram inequivocamente que, de 
fato, a simetria sob tranforma\c c\~oes de paridade era realmente violada
(\nao apenas um pouco mas de maneira m\'axima) nas intera\c c\~oes fracas.

Antes de Lee e Yang assumia-se como evidente que as leis da natureza deveriam
ser sim\'etricas em rela\c c\~ao \`a reflex\~ao em um espelho.
(Pauli tinha rejeitado a equa\c c\~ao para part\'\i culas de spin 1/2 proposta 
por H. Weyl em 1929 com a argumenta\c c\~ao que violava a simetria sob 
paridade.) Por outro lado as intera\c c\~oes eletromagn\'eticas e 
gravitacionais s\~ao invariantes por paridade. A raz\~ao deste fen\^omeno 
ainda \'e um mist\'erio do ponto de vista te\'orico. (Existem algumas propostas 
te\'oricas ainda n\~ao confirmadas.)

Podemos, ent\~ao, dizer que as experi\^encias que em 1957 descobriram
a viola\c c\~ao da paridade foram cruciais no sentido mencionado acima. No 
entanto, \'e interessante notar, que em 1928-30 Richard T. Cox e colaboradores 
realizaram uma s\'erie de experi\^encias de dupla difra\c c\~ao de el\'etrons 
nas quais aparecia uma assimetria segundo a polariza\c c\~ao dos el\'etrons. Agora 
sabemos que isso \'e devido \`a vi\-o\-la\-\c c\~ao da paridade: h\'a uma 
interfer\^encia entre as intera\coes eletromagn\'etica e fraca dos eletrons 
com os n\'ucleons. Por\'em, em 1928 ningu\'em interpretou esses resultados 
desta maneira. Assim, o fato de que as intera\c c\~oes respons\'aveis pelo 
decaimento-$\beta$ n\~ao s\~ao invariantes sob paridade poderia ter sido 
conhecido 27 anos antes de Lee e Yang~\cite{paridade1,paridade2}.

\section{Preconceitos te\'oricos adiantados}
\label{sec:sec3}

Com rela\c c\~ao \`as experi\^encias do tipo {\it ii)} mencionadas na 
Sec.~\ref{sec:intro}, isto \'e, aquelas em que 
preconceitos te\'oricos adiantaram a sua confirma\c c\~ao, podemos citar
quatro:

\noindent {\bf a)} As experi\^encias de E\"otvos para medir a 
rela\c c\~ao entre a massa gra\-vi\-ta\-ci\-o\-nal e a massa inercial. Na 
\'epoca foram aceitos seus resultados. 
Contudo, nos anos 80 um grupo de f\'\i sicos acreditou ter detectado o efeito 
de uma poss\'\i vel ``quinta for\c ca''  com um alcance de 
alguns me\-tros. 
Motivados por isto, os dados das experi\^encias de E\"otvos foram reanalizados 
e observou-se que seus resultados eram compat\ii veis com uma depend\^encia na 
composi\cao qu\ii mica do material (Ver Tabela 1). Assim, 
podemos dizer que, aparentemente,~\footnote{Isto precisa ser melhor estudado.}
a aceita\c c\~ao dos resultados de E\"otvos foi devida 
a preconceitos te\'oricos. A hip\'otese da igualdade entre os dois tipos de 
massa permitiu a formula\c c\~ao da teoria geral da gravita\c c\~ao de 
Einstein. A possibilidade de uma quinta for\c ca foi posteriormente descartada 
por outras experi\^encias~\cite{af1,ef}.

Por outro lado, na d\'ecada dos anos 60, experimentos do tipo de E\"otvos foram
realizados por Robert Dicke e colaboradores. Estes confirmaram a igualdade entre a 
massa gravitacional e a inercial. Contudo, devemos ressaltar que as experi\^
encias de Dicke e E\"otvos  s\~ao sens\'\i veis a (poss\'\i veis) 
for\c cas de alcances diferentes. As \'ultimas mediram efeitos t\'\i picos de 
dimens\~oes de laborat\'orio, as primeiras, t\'\i picos da dist\^ancia Terra-Sol.
Ficam em aberto possibilidades entre dist\^ancias maiores que as de 
laborat\'orios, por\'em menores que a dist\^ancia Terra-Sol ou ainda dist\^ancias 
intergal\'acticas ou mesmo cosmol\'ogicas.

\noindent {\bf b)} A experi\^encia de Millikan sobre a quantiza\c c\~ao da carga
el\'etrica. Ehrenhaft (f\'\i sico alem\~ao) obtinha, com experimentos 
semelhantes, resultados opostos aos de Millikan.
Gerald Holton, revisando as notas de laborat\'orio de Millikan 
observou que, para chegar a seu conhecido resultado, ele eliminou algumas 
medidas que segundo a sua intui\c c\~ao n\~ao eram boas. Caso ele as tivesse 
in\-clu\-\'\i\-do teria chegado \`as mesmas conclus\~oes que Ehrenhaft~\cite{holton1}. 
Historicamente, a hip\'otese de que as cargas el\'etricas possam existir apenas 
em m\'ultiplos inteiros da carga do el\'etron foi muito proveitosa. Isto \'e, 
posteriormente verificou-se que era Millikan quem estava certo.

\noindent {\bf c)} No final da d\'ecada dos anos 20, Hubble mediu a rela\c c\~ao
entre a velocidade de afastamento das gal\'axias em rela\c c\~ao \`a Terra
e sua dist\^ancia at\'e esta. Os dados originais de Hubble est\~ao longe de
serem convincentes~\cite{sw1}. De qualquer forma, assum\'\i-los como corretos
permitiu o modelo cosmol\'ogico, hoje conhecido como {\it modelo
cosmol\'ogico padr\~ao} ou simplesmente como {\it Big Bang}.

\noindent {\bf d)} Finalmente mencionamos a c\'elebre medi\c c\~ao, por uma 
expedi\c c\~ao inglesa, do desv\'\i o da luz pelo Sol prevista por Einstein
segundo a sua teoria geral da relatividade. Hoje sabe-se que as medi\c c\~oes 
de 1919 n\~ao tinham a precis\~ao suficiente. Pior, experi\^encias em 
d\'ecadas posteriores n\~ao confirmaram o valor predito pela relatividade 
geral. Apenas nos anos 70, com a ajuda de sat\'elites, foi poss\'\i vel 
confirmar as predi\c c\~oes te\'oricas~\cite{sw2,jb1}. As duas expedi\c c\~oes
organizadas por Eddington obtiveram os resultados seguintes: a de Sobral, no 
Brasil, $1.98''\pm0.12''$ e a da Ilha de Pr\'\i ncipe $1.61''\pm0.30''$ sendo 
que a teoria de Einstein previa $1.87''$ e a de Newton $0.87''$. No entanto, 
expedi\c c\~oes posteriores n\~ao confirmaram estes resultados. Ver Tabela 2.

Desde 1859 sabia-se da discrep\^ancia na \'orbita de Merc\'urio com a teoria
de Newton. Devido \`a presen\c ca dos planetas (um sistema de
muitos corpos), todas as \'orbitas planet\'arias precessam, isto \'e, a 
orienta\c c\~ao das elipses move-se lentamente no espa\c co. No caso da
\'orbita de Merc\'urio a discrep\^ancia era de 43 segundos de arco por s\'eculo
(575 segundos de arco por s\'eculo contra 532 da teoria Newtoniana). 
\'E interessante que mesmo assim a teoria de Newton n\~ao era considerada 
errada. Havia tamb\'em outras discrep\^ancias, por exemplo a do movimento
da Lua e dos cometas Halley e Encke~\cite{sw2}. Qual delas seria uma 
discrep\^ancia importante? Posteriormente ficou esclarecido que levar em 
conta a press\~ao exercida pelo escape dos gases quando os cometas s\~ao 
aquecidos ao passar perto do Sol, d\'a conta dos movimentos daqueles cometas; e 
que por outro lado devido a Lua ser um corpo extenso, sofre o efeito de 
muitas for\c cas de mar\'e~\cite{sw2}. 
N\~ao foi esse o caso do problema da \'orbita de Merc\'urio como \'e bem conhecido.
Mas \'e curioso que mesmo que a teoria da relatividade geral explicasse esse
problema, foi o desvio da luz pelo Sol que foi considerado como teste 
definitivo dessa teoria. 

Mesmo sem entrar nos detalhes, devemos enfatizar que assumir como corretos
os resultados das quatro experi\^encias anteriores revelou-se importante
do ponto de vista te\'orico e experimental, permitindo assim o desenvolvimento 
posterior das teorias das intera\coes eletromagn\'etica e gravitacional. No 
caso da experi\^encia de Millikan, ela permitiu o desenvolvimento da 
qu\'\i mica como a conhecemos hoje, dado que, a qu\'\i mica com cargas 
fracion\'arias \'e bem di\-fe\-ren\-te daquela com cargas que \sao m\'ultiplos 
inteiros da carga do el\'etron. Contudo, fica claro que crit\'erios alheios ao
m\'etodo cient\'\i fico foram muito importantes.

\section{Experi\^encias no contexto te\'orico errado}
\label{sec:sec4}

Como exemplo das experi\^encias do terceiro tipo mencionamos apenas
duas. Elas mostram que as experi\^encias podem ser guiadas mas n\~ao 
determinadas por teorias particulares.

\noindent {\bf a)}  Os raios c\'osmicos foram descobertos em 1912 por Victor
Hess. Con\-tu\-do, o fato de que est\~ao constitu\'\i dos principalmente por 
pr\'otons, part\'\i culas alfa e n\'ucleos pesados foi confirmado apenas 30 
anos depois. Por muitos anos aceitou-se a hip\'otese de Millikan de que eles eram raios 
gama, isto \'e, radia\c c\~ao e n\~ao part\'\i culas. Em 1929 Bothe e 
Kolh\"oster realizaram uma experi\^encia que confirmou o car\'ater corpuscular 
dos raios c\'osmicos. Eles encontraram que esses raios atravessavam 4.1 cm 
de chumbo. A radia\c c\~ao seria absorvida por essa quantidade de chumbo, assim 
os raios c\'osmicos prim\'arios deviam ser part\'\i culas carregadas. 
Mesmo que essa conclus\~ao esteja correta, o que Bothe e Kolh\"oster observaram
n\~ao foram os raios c\'osmicos prim\'arios, mas os secund\'arios: eram os muons que 
ainda n\~ao tinham sido descobertos~\cite{ot,lon}!

\noindent {\bf b)} A experi\^encia de Stern-Gerlach realizada em 1921, 
demostrou a ``quantiza\c c\~ao espacial'' ou seja a quantiza\c c\~ao das 
orienta\c c\~oes permitidas do momento angular orbital dos \'atomos. De 
fato, esta experi\^encia foi proposta para verificar ou a teoria cl\'assica do 
momento angular de Larmor ou a teoria qu\^antica de Bohr-Sommerfeld. As duas 
teorias faziam predi\c c\~oes bem definidas do comportamento dos \'atomos em 
campos magn\'eticos uniformes. A teoria cl\'assica e a de Bohr-Sommerfeld 
diferiam nas suas predi\c c\~oes dos valores do momento magn\'etico $\vec\mu_l$ 
e a distribui\c c\~ao de m\'aximos e m\'\i nimos observados na tela depois que 
os \'atomos atravessavam um campo magn\'etico n\~ao homog\^eneo. Otto Stern e
Walter Gerlach mediram os valores poss\'\i veis
para a componente $\mu_{lz}$ para \'atomos de prata~\cite{cohent}. 

A teoria cl\'assica permite que $\mu_{lz}$ possa ter qualquer valor entre
$-\mu_{lz}$ e $+\mu_{lz}$. O resultado da experi\^encia foi:
\begin{equation}
\label{e1}\mu_{lz}=\pm\,\frac{e\hbar}{2m}\,.
\end{equation}
Isto \'e, o feixe de \'atomos de prata \'e separado em duas componentes
discretas, uma no sentido positivo do eixo-$z$ e outra no sentido negativo.
(O resultado \'e independente da escolha do eixo.) 

A teoria de Bohr-Sommerfeld previa a quantiza\c c\~ao dos planos orbitais
dos el\'etrons at\^omicos. Quer dizer que, al\'em da proposta de Bohr segundo 
a qual as elipses dos el\'etrons ao redor do n\'ucleo tinham o seu tamanho e 
forma quantizada, segundo Sommerfeld, tamb\'em a orienta\c c\~ao da \'orbita 
em rela\c c\~ao a um eixo apropriado, por exemplo o determinado por um campo
magn\'etico, poderia a ter apenas valores discretos. Al\'em do n\'umero 
qu\^antico principal (radial) $n'$, Sommerfeld introduziu um n\'umero 
qu\^antico azimutal $n$, com $n=n_1+n_2$ e o n\'umero permitido de planos das 
\'orbitas seria $2n$ incluindo as posi\c c\~oes horizontais e verticais, isto 
\'e, par. (Na nota\c c\~ao atual $n_1=m_l$, e $n_2=l$, sendo $m_l$ o n\'umero 
qu\^antico azimutal e $l$ o n\'umero qu\^antico 
para o momento angular orbital.)  Stern tinha em mente este contexto te\'orico 
quando realizou a famosa experi\^encia.

Na teoria de Bohr-Sommerfeld o caso $n=0$~\cite{fw} era proibido como veremos 
mais adiante mas neste caso 
o feixe de \'atomos n\~ao seria desviado. 
Como eles imaginavam que estavam no caso $n=1$ ($l=1$ na nota\c c\~ao
a\-tu\-al) deveriam 
esperar duas componentes na teoria de Bohr-Sommerfeld. Na verdade eles estavam 
no caso $l=0$ e, na mec\^anica qu\^antica de Schr\"odinger se esperaria que o 
feixe n\~ao fosse desviado. O caso $l=1$ produziria tr\^es componentes mesmo 
no contexto da mec\^anica qu\^antica de Schr\"odinger dado que a 
degeneresc\^encia \'e dada por $2l+1$.

Stern e Gerlach n\~ao sabiam que a teoria de Bohr-Sommerfeld seria em breve 
substitu\'\i da pela mec\^anica qu\^antica e que al\'em disso, existe uma outra
fonte de momento angular diferente da que tinha sido considerada
at\'e ent\~ao. Depois da proposta da exist\^encia do 
{\it spin} do el\'etron, Phipps e Taylor repetiram a experi\^encia de Stern e 
Gerlach, por\'em usando desta vez \'atomos de hidrog\^enio. Eles obtiveram os 
mesmos resultados que Stern-Gerlach, mas n\~ao os
associaram com o spin. Como os \'atomos de hidrog\^enio t\^em mais 
pro\-ba\-bi\-li\-da\-de 
de estar no estado fundamental, $l=0$, sem o spin n\ao deveria haver desvio do 
feixe~\cite{pt}. Agora sabemos, de fato, que o efeito 
Stern-Gerlach \'e devido \`a exist\^encia do spin do el\'etron. Tamb\'em \'e 
de notar que Uhlenbeck e Goudsmith (que propuseram o spin do el\'etron em 
1925) tampouco notaram que a experi\^encia de Stern-Gerlach era uma 
evid\^encia do novo grau de liberdade introduzido por eles~\cite{fw}.

\section{Duas experi\^encias cruciais: A viola\c c\~ao da paridade e a
 quantiza\c c\~ao espacial}
\label{sec:5}
Os problemas da viola\c c\~ao da paridade e da quantiza\c c\~ao espacial,
comentados brevemente nas Secs.~\ref{sec:sec2} e \ref{sec:sec4}, 
respectivamente, podem ser considerados {\it cruciais} no sentido da 
experi\^encia de Newton mencionada no Se\c c\~ao~\ref{sec:intro}. Permitiram 
descartar teorias que conservam a paridade, no primeiro caso; e a 
mec\^anica cl\'assica, no segundo caso. No entanto, n\~ao podemos dizer 
que eles confirmaram uma teoria em particular.
Estas duas experi\^encias apresentam caracter\ii sticas especiais que merecem
ser analisadas.

Voltemos ao caso da viola\c c\~ao da paridade. 
Em 1926 Davisson e Germer~\cite{dg1} realizaram a experi\^encia que mostrou
que a hip\'otese de de Broglie e Schr\"odinger sobre a natureza ondulat\'oria 
das part\'\i culas estava correta. Isso incentivou, segundo as pr\'oprias 
palavras de Cox~\cite{paridade2}, um estudo mais detalhado da natureza 
dessas ondas: s\~ao longitudinais, transversais ou t\^em ambas 
carater\'\i sticas?~\footnote{Na \'epoca ainda se pensava na 
exist\^encia real dessas ondas e n\~ao em termos de pro\-ba\-bi\-li\-da\-des 
como prop\^os pouco tempo depois Max Born.}  
Foi com essa motiva\c c\~ao te\'orica
que em 1928 Charles G. McIlwrait, Bernard Kurrelmeyer e Cox come\c caram uma 
experi\^encia na Universidade de Nova York~\cite{cox1}. Eles esperavam que, 
como no caso dos raios-X que s\~ao polarizados via duplo espalhamento, o mesmo 
acontecesse com os el\'etrons caso eles fossem ondas transversais. Eles 
acreditavam 
que o resultado seria positivo pois na \'epoca j\'a havia sido proposto o 
{\it spin} do el\'etron por Goudsmit e Uhlenbeck. O curioso (e Cox n\~ao lembra 
o motivo) \'e que em vez de usarem como fonte dos el\'etrons um filamento 
quente, como era usual na \'epoca, eles usaram el\'etrons produzidos em 
decaimentos radiativos (produzidos, agora sabemos, pelas 
intera\c c\~oes fracas!).
Afinal, porque deveriam ser diferentes os dois tipos de el\'etrons?
Eles n\~ao ti\-nham ouvido falar da conserva\c c\~ao da paridade. Devemos 
lembrar que em 1927 Wigner tinha explicado as leis de sele\c c\~ao de 
Laporte em transi\c c\~oes at\^omicas invocando a conserva\c c\~ao da paridade. 
Esta simetria \'e, de fato, conservada nas intera\c c\~oes eletromagn\'eticas.
Esta foi a primeira vez que essa simetria foi usada na explica\c c\~ao
das leis da f\'\i sica. Antes disso era usada apenas como recurso para resolver 
problemas com maior facilidade.

Na experi\^encia de Cox, os el\'etrons eram espalhados duas vezes a 90$^0$
em alvos de ouro. Observaram uma assimetria: menos part\'\i culas $\beta$ foram
observadas a 90$^0$ do que a 270$^0$. Depois de longa an\'alise
conclu\'\i ram que se tratava da polariza\c c\~ao dos el\'etrons-$\beta$ vindo
da fonte de r\'adio (Ra). A sua
d\'uvida nesta interpreta\c c\~ao \'e manifesta no t\'\i tulo do seu artigo
``Apparent evidence of polarization in a beam of $\beta$-rays''~\cite{cox1}.
Essa assimetria, conjecturaram, aparecia apenas nos el\'etrons de alta energia!
Segundo a teoria $V-A$ das intera\c c\~oes fracas proposta em 1958, a 
polariza\c c\~ao dos el\'etrons produzidos em decaimentos $\beta$ \'e dada por 
$P=\pm v/c$. Se $v\rightarrow0$ ent\~ao $P\rightarrow 0$ {\it i.e.}, 
n\~ao temos polariza\c c\~ao, por\'em se $v\rightarrow c$ temos que 
$P\rightarrow \pm1$, a massa dos el\'etrons \'e neste caso desprez\'\i vel e eles 
se comportam como neutrinos.

Em 1929 Carl T. Chase realizou tr\^es experi\^encias semelhantes \`a
de Cox {\it et al.}, desta vez usando como fonte dos el\'etrons-$\beta$ uma 
amostra de rad\^onio (Rn). Na primeira n\~ao apareceu a 
assimetria mas os el\'etrons eram de baixa energia. Na segunda experi\^encia 
Chase confirmou que, como Cox {\it et al.,} tinham
conjecturado, os contadores eram sens\'\i veis \`a velocidade dos el\'etrons.
A assimetria desaparecia para el\'etrons de baixa velocidade. Na terceira
experi\^encia, o contador Geiger, usado at\'e ent\~ao em todas as experi\^encias,
foi substituido por um eletrosc\'opio muito sens\ii vel e a assimetria, mesmo
pequena, apareceu de maneira consistente.

Uma quinta experi\^encia foi feita por Frank E. Myers e Cox para estudar a 
assimetria em el\'etrons-$\beta$. Desta vez, em analogia com a polariza\c c\~ao
da luz por um cristal de turmalina, fizeram uma contagem dos el\'etrons que
passavam atrav\'es de duas l\'aminas finas de ferro magnetizadas em diferentes 
dire\c c\~oes. N\~ao encontraram diferen\c ca na contagem quando as 
dire\c c\~oes das magnetiza\c c\~oes eram mudadas. 

Chase e Myers e outros pesquisadores da universidade de Nova York continuaram
as experi\^encias relacionadas com a polariza\c c\~ao de  el\'etrons r\'apidos
mas usando filamentos quentes como fonte dos el\'etrons depois acelerados
por alta voltagem. N\~ao encontraram nenhuma assimetria obviamente, para n\'os 
agora, porque eram eletrons n\ao polarizados. Segundo 
Cox~\cite{cox1}
\begin{quotation}
``This later work of Chase, Myers and others left me quite puzz\-led.
I found it difficult to reconcile our observations of $\beta$-particles with
the prevailing theory and observations on artificially accelerated
electrons. I never published a retraction of our findings, but I probably
express some doubts about the reality of our effect in conversation with
friends''.
\end{quotation}

Em 1929 Mott mostrou que a intera\c c\~ao spin-\'orbita no espalhamento de
Coulomb pode ser usada como polarizador via duplo
espalhamento para el\'etrons relativ\'\i sticos e,
segundo Lee Grodzins as experi\^encias de Cox e colaboradores n\~ao eram
apropriadas para observar o efeito Mott. As amostras eram grossas, assim
o espalhamento n\~ao era duplo mas m\'ultiplo. 

Em 1960, foi realizada no MIT  uma experi\^encia n\~ao publicada (por Sidney 
Altman como sua Senior Thesis) que mostrava que o sinal da assimetria de Cox 
{\it et al.}, estava errado~\cite{lee}. O sinal da assimetria da 
experi\^encia de Cox
{\it et al.,} foi explicado depois como um erro n\~ao da experi\^encia em si
mas na escolha das coordenadas~\cite{paridade2}.

Independente disso, podemo
perguntar-nos por que n\~ao foi reconhecido na \'epoca, que a assimetria de Cox
{\it et al.}, corresponde \`a viola\c c\~ao da paridade. Nessa \'epoca houve
muita atividade experimental para detectar o efeito Mott. Assim, as 
experi\^encias de Cox podem-se ter confundido com outras similares mas que 
procuravam o efeito Mott~\cite{paridade2}. As experi\^encias de 1928-1930 
mostraram uma polariza\c c\~ao longitudinal para o el\'etron e isso significa 
uma viola\c c\~ao da paridade. Ela mostrava um n\'umero de eventos a 90$^0$ 
diferente dos observados a 270$^0$. O espalhamento Mott n\~ao previa uma 
assimetria nesses \^angulos mas entre 0$^0$ e 180$^0$. 
A polariza\c c\~ao de Mott foi detectada experimentalmente apenas em 
1942~\cite{hat}. (Em 1935 Halpern e Schwinger observaram que a n\ao observa\cao 
do efeito Mott poderia ser devida ao que agora chamamos de ``corre\coes 
radiativas'' ou a desvios do campo nuclear de sua forma Coulombiana~\cite{js}.)

Por outro lado, as experi\^encias
de Wu {\it et al.} e outros (ver embaixo), podem ser consideradas como 
cruciais porque elas foram capazes de distinguir entre dois tipos de teorias: 
as que conservam a paridade e aquelas que a violam~\cite{paridade3}. Por que
isso n\~ao foi reconhecido em 1928 e 1930? 

Qual foi o contexto em 1956 quando Lee e Yang propuseram que a viola\c c\~ao da 
paridade poderia acontecer nas intera\c c\~oes fracas?

Em 1950 Purcell e Ramsey haviam colocado a quest\~ao
da viola\c c\~ao da paridade no contexto do momento dipolar 
el\'etrico~\cite{pr}. Tamb\'em Wick, Whitman e Wigner em seu famoso artigo de
1952~\cite{www} con\-si\-de\-ra\-ram o conceito da paridade intr\ii nseca das 
part\'\i culas elementares. Eles observaram que
a paridade poderia ser uma simetria aproximada, da mesma maneira que a simetria
que transforma mat\'eria em antimat\'eria (e viceversa) e, apenas as duas
transforma\c c\~oes combinadas seria uma simetria da natureza. Estas 
duas considera\c c\~oes da viola\c c\~ao da paridade foram apenas uma forma
especulativa e n\~ao propunham esta como solu\c c\~ao de um problema 
determinado.

Em 1953 foi colocado {\it experimentalmente} o chamado
``paradoxo $\tau-\theta$''. Duas part\'\i culas a $\tau$
e a $\theta$ tinham a mesma massa, spin e vida m\'edia mas diferentes canais 
de decaimento.~\footnote{A $\tau$ n\~ao \'e 
a part\'\i cula que agora conhecemos como lepton $\tau$. Na nota\c c\~ao atual
ambas $\tau$ e $\theta$ correspondem ao kaon $K^+$.} 
A diferen\c ca estava no fato de
que a paridade nos decaimentos era oposta em cada caso. Assim, se a paridade 
fosse conservada elas teriam que ser part\'\i culas diferentes; se n\~ao o 
fosse, elas poderiam ser a mesma part\'\i cula. Lee e Yang, analisando todas as 
experi\^encias dos decaimentos nucleares $\beta$ conhecidas na \'epoca 
observaram que todas eram desenhadas de forma tal que mesmo que a paridade 
fosse violada nesse decaimento, as experi\^encias feitas at\'e ent\~ao n\~ao 
a teriam detectada. Propuseram ent\~ao que se fizessem 
v\'arios testes para verificar se a simetria sob paridade \'e ou n\~ao 
conservada nesses processos fracos. Entre eles estavam o decaimento $\beta$ 
de n\'ucleos com spin orientado. Se $\phi$ \'e o \^angulo entre o spin do 
n\'ucleo que decai e o momento do el\'etron, esperava-se, caso a paridade 
fosse violada, uma assimetria
na distribui\c c\~ao angular entre $\phi$ e $\phi-180^0$. Um segundo tipo
de assimetria apareceria na distribui\c c\~ao angular na cadeia 
$\pi\to \mu\nu$, $\mu\to e\nu\nu$. Ambas experi\^encias foram realizadas nos 
seis meses seguintes \`a proposta de Lee e Yang por C. S. Wu, E. Ambler, R. W. 
Hayward, D. D. Hoppes e R. P. Hudson no primeiro caso;  e por  R. L. Garwin, 
L. M. Lederman e M. Weinrich e, independentemente, por  J. I. Friedman e V. L. 
Telegdi, no segundo caso~\cite{cox1}.
Assim vemos na quest\~ao da paridade que o conte\'udo f\'\i sico e l\'ogico 
da experi\^encia n\~ao \'e suficiente para revel\'a-la como crucial. Mas, se o 
contexto te\'orico \'e importante, por que n\~ao o foi no caso da 
quantiza\c c\~ao espacial? 

No caso da experi\^encia de Stern-Gerlach a conclus\~ao correta foi que
o aparecimento de duas ``manchas'', em vez de um cont\'\i nuo na tela, refutava
a teoria cl\'assica. Por\'em, como mencionamos antes, ao mesmo tempo cri\-a\-va
dificuldades para a teoria de Bohr-Sommerfeld. Quando considerada no contexto
de dois tipos de momentos angulares, o {\it momento angular orbital}, $\vec{L}$,
e o {\it momento angular de spin}, $\vec{S}$, a experi\^encia de Stern-Gerlach 
\'e uma refuta\cao da velha mec\^anica qu\^antica. Como $l=0$ nos \'atomos de 
prata no estado fundamental, a separa\cao do feixe em duas partes observada por
Stern-Gerlach foi devida ao momento angular $\vec{S}$, que na \'epoca ainda n\ao
tinha sido proposto~\cite{fw}. Stern e Gerlach achavam que estavam no caso
$l=1$ e segundo a teoria de Bohr-Sommerfeld esperavam $2l$ componentes.
Mas na mec\^anica qu\^antica o n\'umero delas \'e $2l+1$; assim deveriam
aparecer tres manchas na tela. Como agora sabemos que o \'atomo de prata no estado 
fundamental est\'a no estado $l=0$, do ponto de vista da moderna mec\^anica 
qu\^antica n\~ao esperamos separa\c c\~ao do feixe a menos que exista uma outra 
fonte de momento angular.
Em breve a velha teoria qu\^antica seria substitu\'\i da pela 
moderna mec\^anica qu\^antica (e o sp\'\i n pela extens\ao relativ\ii stica de 
Dirac).
A raz\~ao da exclus\~ao do caso $n=0$ ($l=0$ na nota\cao atual) na teoria
de Bohr-Sommerfeld \'e porque neste caso as \'orbitas el\'\i pticas degeneram
em linhas retas. Isto implica que o el\'etron colide com o n\'ucleo, o que
\'e observado~\cite{born}.

\section{Conclus\~oes}
\label{sec:con}
Podemos notar de uma maneira muito intuitiva, que as experi\^encias
cient\'\i ficas n\~ao s\~ao, em geral, um meio de falsificar as teorias. 
A experi\^encia de Stern-Gerlach tem certa independ\^encia da teoria: estava 
baseada em contextos te\'oricos errados. Quer dizer que as experi\^encias 
podem ter a sua pr\'opria raz\~ao de ser. Elas podem sugerir novos avan\c cos 
te\'oricos, podem estar baseadas em conceitos te\'oricos falsos e mesmo assim 
produzir resultados verdadeiros. Podem tamb\'em dar evid\^encia contra uma 
teoria concorrente. 
Dos exemplos anteriores podemos ver que devemos fazer uma diferen\c ca entre
uma teoria que permite realizar uma experi\^encia e uma teoria que est\'a 
sendo testada~\cite{fw}.

Assim, se nos perguntarmos: quando uma experi\^encia \'e crucial? devemos 
admitir que n\~ao temos um resposta bem definida.
Normalmente acredita-se, como foi discutido na Sec.~\ref{sec:intro}, 
que as experi\^encias {\it cruciais} s\ao aquelas
que decidiram entre duas teorias rivais. Vimos no caso da quantiza\cao
espacial (Stern-Gerlach) que ela n\~ao foi suficiente para confirmar (pelo 
contr\'ario, era contra) a velha mec\^anica qu\^antica. No entanto, sendo seu 
resultado v\'alido, em breve seria interpretada corretamente na teoria qu\^antica 
moderna com a introdu\cao de um novo grau de liberdade: o {\it sp\'\i n} do 
el\'etron. Por outro lado as experi\^encias de Cox {\it et al.} n\~ao foram 
suficientes para revelar aos f\ii sicos que a simetria por paridade n\ao \'e 
conservada em certos processos (agora conhecidos por intera\coes fracas.)

A experi\^encia de Stern-Gerlach deu um resultado que
estava contra as previs\~oes da mec\^anica cl\'assica, da mec\^anica 
qu\^antica antiga (de Bohr-Sommerfeld) e mesmo contra as da mec\^anica 
qu\^antica de Schr\"odinger (que n\~ao considerava o grau de liberdade de 
spin). Os resultados experimentais estavam corretos mas a sua 
interpreta\c c\~ao te\'orica n\~ao. Assim n\~ao \'e completamente certo que 
uma experi\^encia decisiva pode ser considerada apenas \`a luz de uma 
teoria~\cite{fw}. Uma experi\^encia pode ser guiada por uma ou v\'arias 
teorias mas n\~ao determinada por elas. Uma experi\^encia pode falsificar uma 
teoria (a de Lamor no caso da experi\^encia de Stern-Gerlach) mas n\~ao 
validar uma outra teoria rival (a mec\^anica qu\^antica de Bohr-Sommerfeld). 

Se aceitamos que \'e a comunidade cient\'\i fica quem, num determinado contexto
te\'orico, determina se um resultado experimental \'e crucial ou
n\~ao, ent\~ao teremos que aceitar que as verdades cient\'\i ficas t\^em 
tamb\'em um car\'ater {\sl hist\'orico}. Na \'epoca de Millikan, 
foi importante aceitar os seus resultados e n\~ao os de Erhenhaft.
Da mesma maneira, os de Hubble e E\"otvos. Neste \'ultimo caso \'e interessante
notar que, se a depend\^encia na composi\c c\~ao des\-co\-ber\-ta nos dados de 
E\"otvos tivesse sido observada na \'epoca s\'o poderia ter sido interpretada
como uma diferen\c ca entre a massa gravitacional e a massa inercial: na 
\'epoca o n\^eutron n\~ao tinha sido descoberto!  Na atualidade, caso a 
depend\^encia na composi\c c\~ao tivesse sido confirmada, n\~ao alteraria o 
princ\'\i pio de equival\^encia porque conhecemos as cargas \`as quais atribuir esse 
efeito, por exemplo, o isospin ou o n\'umero bari\^onico~\cite{gm}. Mais uma 
vez vemos que aceitar os resultados de E\"otvos foi historicamente correto.

Mas se o contexto te\'orico 
\'e importante, como explicar o caso dos raios-X e da radioatividade? Estes
foram descobertos sem nenhum marco te\'orico. Cons\-ti\-tu\-em um tipo de 
descobertas completamente inesperadas~\cite{es}. \'E dif\'\i cil de se prever, 
se no futuro isso vai acontecer de novo ou n\~ao. 
Tamb\'em n\~ao da para entender 
por que os resultados de Cox e colaboradores n\~ao es\-ti\-mu\-la\-ram mais 
pesquisas nessa dire\c c\~ao. Davisson e Germer realizaram uma 
experi\^encia semelhante \`a de Cox {\it et al.}, na qual n\~ao observaram 
nenhuma assimetria~\cite{dg2}. Mas eles usaram el\'etrons produzidos por um 
filamento o que  significa uma fonte n\~ao polarizada de el\'etrons. O 
importante de novo, \'e que a comunidade n\~ao percebeu na \'epoca que os 
el\'etrons-$\beta$ eram o produto de uma nova intera\c c\~ao: a 
for\c ca ou intera\cao fraca, como seria conhecida mais tarde.

Vemos que, pelo menos nos exemplos discutidos acima, resultados
experimentais permanecem, at\'e certo ponto, independentes da 
interpreta\c c\~ao te\'orica. Isto d\'a \`as experi\^encias uma relev\^ancia 
ainda maior que a que \'e usualmente aceita. Por\'em, em alguns casos 
antecipamo-nos \`a natureza guiados por um senso est\'etico (mesmo mal definido)
como foi o caso da teoria da relatividade geral e a sua ``confirma\c c\~ao''
por Eddington em 1919. Mas, uma experi\^encia sempre \'e {\it guiada} por
algum contexto te\'orico ainda que este seja errado (como no caso de S-G)
mas, mesmo sendo dirigida teoricamente, algumas experi\^encias n\ao s\ao
{\it determinadas} pela teoria. Ent\~ao, uma teoria n\ao precisa ser correta
para ser experimentalmente importante. A cren\c ca popular entre os
cientistas de que uma experi\^encia somente \'e decisiva
\`a luz de uma teoria \'e um pouco relaxada. 

N\~ao podemos criar o conhecimento cient\'\i fico pela raz\~ao pura como queria
Kant mas, por outro lado, as teorias s\~ao inven\c c\~oes livres do esp\ii rito 
humano (qualquer coisa que isso signifique). Mas essa liberdade n\~ao \'e
completa, ela est\'a limitada pelos dados experimentais. Ningu\'em poderia,
impunemente, propor uma teoria da gravita\c c\~ao cl\'assica ou da 
eletrost\'atica com uma for\c ca dependendo de $1/r^{3}$. As vezes, uma 
formula\c c\~ao te\'orica adianta-se aos dados experimentais, como foi
o caso da teoria geral da relatividade. Por outro lado, mesmo que 
aceitemos que a ci\^encia em geral, e a f\'\i sica em particular, 
est\ao baseadas na experi\^encia h\'a v\'arias maneiras de se interpretar 
isso. Para alguns, como Mach, as cadeias de racioc\ii nio l\'ogico, por 
complicadas que sejam, s\~ao em \'ultima inst\^ancia,
rela\c c\~oes entre {\it fatos}. A teoria seria uma maneira econ\^omica de 
relacionar experi\^encias n\~ao sendo ela mesma um elemento do 
conhecimento~\cite{ac}. A maneira de conciliar esse ponto de vista com as
teorias da relatividade e da mec\^anica qu\^antica foi, para positivistas 
l\'ogicos, aceitar que a ci\^encia permite explica\c c\~oes baseadas em 
conceitos abs\-tra\-tos e princ\'\i pios gerais sempre que sejam relacionados de 
maneira l\'ogica com consequ\^encias {\it verific\'aveis} ou {\it false\'aveis} 
segundo Popper. 

A f\'\i sica, em geral, precisa que uma teoria preceda uma experi\^encia 
porque sem uma teoria n\~ao \'e poss\'\i vel extrair os fatores que s\~ao 
relevantes.
Sem um contexto te\'orico apropriado, como poderia William Crooker ter 
descoberto os raios-X, quando observou que as placas fotogr\'aficas de seu 
labor\'atorio estavam estragadas? (ele as devolveu aos fabricantes!) ou Cox 
reconhecer que a simetria por paridade \'e violada em alguns processos que 
envolvem el\'etrons? Mas Roentgen, nas mesmas circunst\^ancias de Crookes e 
outros, foi capaz de demostrar que um novo tipo de radia\c c\~ao (diferente da 
luz vis\ii vel) existia. Apenas ci\^encias nos seus prim\'ordios podem realizar 
experi\^encias sem um contexto te\'orico bem determinado. Veja-se por exemplo, 
os trabalhos sobre a m\'aquina de v\'acuo de Boyle e as primeiras pesquisas 
sobre termodin\^amica.

Na atualidade, o caso da f\ii sica de altas energias est\'a numa fase em que
as experi\^encias precisam, talvez mais do que nunca, uma teoria para serem
interpretadas. Este tema merece um artigo separado.

Talvez, como prop\~oe Weinert, seja necess\'ario a constru\cao de uma 
{\it epistemologia experimental}~\cite{fw}.

\vskip .5cm

Agrade\c co a A. Susuki pela leitura cr\'\i tica do manuscrito e ao CNPq
pelo apoio financeiro parcial

\eject
\begin{table}
\begin{center}
\begin{tabular}{||l|l||} \hline
Subst\^ancia & $(x-x_{pt})\times 10^{-6}$ \\ \hline
magnesita & $+0.004\pm0.001$ \\
madeira & $-0.001\pm0.002$ \\
cobre   & $+0.004\pm0.002$ \\
agua    & $-0.006\pm0.003$ \\
sulfato de cobre (cristalino) & $-0.001\pm0.003$ \\
sulfato de cobre (solu\c c\~ao) & $-0.003\pm0.003$ \\
asbestos & $+0.001\pm0.003$ \\
sebo & $-0.002\pm0.003$ \\ \hline
\end{tabular}
\caption{Dados da experi\^encia de E\"otvos et al. Ref.~[FR93].}
\end{center}
\end{table}

\begin{table}
\begin{center}
\begin{tabular}{||l|l|l|l||} \hline 
Observat\'orio & lugar      & data           & resultado $\pm$ erro \\ \hline
Greengrich     & Australia  & set. 21, 1922  & $1.77\pm0.40$ \\
Postdam        & Sumatra    & maio 9, 1929   & $2.24\pm0.10$ \\
Sternberg      & USSR       & junho 19, 1936 & $2.73\pm0.31$ \\
Yerkes         & Brasil     & maio 20, 1947  & $2.01\pm0.27$ \\
Yerkes         & Sudan      & fev. 25  1952  & $1.70\pm0.10$ \\
Texas          & Maurit\^ania & junho 30, 1973 & $(0.94\pm0.11)\times L_E$\\ 
\hline
\end{tabular}
\end{center}
\caption{Resultados de experi\^encias sobre o desvio da luz eplo sol Ref.~[BE93].
O valor predito pela relatividade geral \'e $L_E=1.74$.}
\end{table}
\end{document}